\documentclass[lettersize,journal]{IEEEtran}
\usepackage{cite}
\usepackage{amsmath,amssymb,amsfonts}
\usepackage{algorithmic}
\usepackage[ruled,vlined]{algorithm2e}
\usepackage{graphicx}
\usepackage{textcomp}
\usepackage{bm}
\usepackage[ruled]{algorithm2e}
\usepackage{booktabs}
\usepackage{multirow}
\usepackage{xcolor}
\usepackage{makecell}

\begin{document}
\title{Tomographic Foundation Model---FORCE: Flow-Oriented Reconstruction Conditioning Engine}

\author{Wenjun Xia, Chuang Niu, and Ge Wang
\thanks{W. Xia, C. Niu, and G. Wang are with the Department of Biomedical Engineering, School	of Engineering, Rensselaer Polytechnic Institute, Troy, 12180, NY, USA (e-mails: xiaw4@rpi.edu, niuc@rpi.edu, wangg6@rpi.edu). }
}
\maketitle

\begin{abstract}
Computed tomography (CT) is a major medical imaging modality. Clinical CT scenarios, such as low-dose screening, sparse-view scanning, and metal implants, often lead to severe noise and artifacts in reconstructed images, requiring improved reconstruction techniques. The introduction of deep
learning has significantly advanced CT image reconstruction. However, obtaining paired training data remains rather challenging due to patient motion and other constraints. Although deep learning methods can still perform well with approximately paired data, they inherently carry the risk of hallucination due to data inconsistencies and model instability. In this paper, we integrate the data fidelity with the state-of-the-art generative AI model, referred to as the Poisson flow generative model (PFGM) with a generalized version PFGM++, and propose a novel CT framework: Flow-Oriented Reconstruction Conditioning Engine (FORCE). In our experiments, the proposed method shows superior performance in various CT imaging tasks, outperforming existing unsupervised reconstruction approaches.
\end{abstract}

\begin{IEEEkeywords}
CT reconstruction, Poisson flow generative model, unsupervised learning, iteratively reconstruction
\end{IEEEkeywords}

\section{Introduction}
\label{sec:1}
Computed tomography (CT) is a widely used imaging modality in clinical practice, homeland security, industrial evaluation, and other domains. In 2023 alone, approximately 93 million CT scans were performed in 62 million patients in the United States, a number that continues to grow \cite{smith2025projected}. However, the ionizing radiation associated with CT remains a major health concern. Arguably, CT-related exposure could contribute to approximately 5\% of newly diagnosed cancers annually, a total of estimated 103,000 radiation-induced cancer cases in 2023 \cite{smith2025projected}. Hence, reducing CT radiation dose has become a global priority, as evidenced by the ”As low as reasonably achievable” (ALARA) guideline.

Low-dose and sparse-view scanning are two common strategies to reduce radiation exposure to a patient. In principle, these approaches would compromise image quality, leading to increased image noise and artifacts that can hinder accurate diagnosis. To balance radiation reduction with diagnostic performance, advanced denoising and artifact removal techniques are essential. In addition, metal implants are frequently encountered in clinical settings. These high-density materials absorb nearly all incident X-ray photons, resulting in severe metal artifacts and information loss. Since the invention of CT in 1972, imaging researchers have been developing robust CT reconstruction algorithms to address these critical challenges.

Filtered back-projection (FBP) \cite{kak2001principles} remains the most classic and widely used algorithm for CT image reconstruction. Although computationally efficient, FBP is highly sensitive
to noise and incompleteness of data, making it inadequate for handling corrupted measurements. To address these limitations, iterative reconstruction (IR) algorithms were developed, including the algebraic reconstruction technique (ART) \cite{gordon1970algebraic}, simultaneous iterative reconstruction technique (SIRT) \cite{trampert1990simultaneous}, simultaneous algebraic reconstruction technique (SART) \cite{andersen1984simultaneous} and  expectation maximization (EM) \cite{dempster1977maximum}. These methods formulate CT reconstruction as a linear inverse problem and iteratively solve a system of linear equations, offering improved noise and artifact suppression. However, these IR methods still struggle to effectively address the severe image degradation caused by a low signal-to-noise ratio of data, a sparse sampling pattern, and the presence of metal implants in a patient body.

Over the past few decades, CT reconstruction methods have evolved from traditional algorithmic designs to deep learningbased approaches, with deep tomographic reconstruction becoming the mainstream in the field.

Traditional methods primarily aim to suppress noise and artifacts through classic techniques such as filtering, iterative correction, and regularization, applied in projection and/or image domains. For example, Balda et al. \cite{balda2012ray} and Manduca et al. \cite{manduca2009projection} filtered projection data for noise reduction. 
In metal artifact reduction (MAR), many researchers interpolated high-quality data to estimate corrupted data in metal-affected regions \cite{lewitt1978image, kalender1987reduction}. Meyer et al. \cite{meyer2010normalized} further refined this approach by segmenting different tissue types, enabling smoother interpolation and less residual artifacts. In addition, sparsity-based optimization in the projection domain was introduced to enforce smoother data consistency, enhancing denoising performance \cite{wang2006penalized, karimi2016sinogram} and MAR \cite{duan2008metal, xue2009metal, zhang2011new}.

Among traditional methods, model-based iterative reconstruction (MBIR) is a central focus. Rooted in Bayesian statistical theory, MBIR formulates CT reconstruction as an optimization problem by combining a data fidelity term derived from the physical CT measurement model with a prior term that encodes the assumption or prior knowledge about underlying images. This prior as the regularization term promotes desirable image properties. The solution is then obtained through iterative optimization. Over time, MBIR enjoyed substantial development, with numerous priors including total
variation (TV) and its higher-order extensions \cite{yu2005total, sidky2008image, zhang2013few, niu2014sparse, zhang2014few, wu2018low, wu2020high}, sparsity and low-rank models \cite{xu2012low, kim2014sparse, bao2019convolutional, zhang2016spectral, chun2017sparse, xia2019spectral, chen2022font}, and many others \cite{ chen2009bayesian, wang1996iterative, ma2012iterative}.
Despite its substantial progress and application, MBIR remains constrained by major efforts in crafting prior models and tuning hyper-parameters, as well as high computational overheads in clinical tasks.

Over the past years, deep learning has rapidly emerged as a powerful alternative, offering data-driven solutions capable of learning complex mappings and image priors directly from large-scale datasets. As a result, deep learning has advanced the field of medical imaging \cite{wang2016perspective, wang2019machine}. Current deep learning-based CT reconstruction methods can be broadly categorized into three categories:

\textbf{(1) Sinogram-domain methods} operate directly on projection data to suppress noise or recover missing data. Some approaches focus on denoising raw sinograms before reconstruction \cite{yang2022low, ma2021sinogram}, while others aim to restore data corrupted by metal parts or missed by sparse sampling using learned completion models \cite{park2018ct, ghani2018deep, ghani2019fast}.

\textbf{(2) Image-domain methods} are widely adopted due to their effectiveness in artifact removal and image enhancement, without the need to access raw data. These methods learn mappings from corrupted images to artifact-free counterparts \cite{chen2017low, chen2017lowdose, jin2017deep, kang2017deep, yang2018low, niu2022noise, liao2019adn, wang2021dicdnet, niu2021low, yu2021metal}. A related line of work involves unrolling MBIR algorithms into deep networks, allowing the network to learn prior knowledge through iterative updates \cite{chen2018learn, gupta2018cnn, he2018optimizing, wang2022adaptive, xiang2021fista, xia2021magic, xia2023regformer}. In some MAR methods, networks are also used to generate prior images, which are subsequently forward-projected for sinogram correction \cite{zhang2018convolutional, gjesteby2019dual, yu2020deep}.

\textbf{(3) Hybrid methods} work in both sinogram and image domains to improve robustness and accuracy of image reconstruction. Some models learn mappings from the projection domain to the image domain using fully-connected or Radon transform-based architectures \cite{zhu2018image, he2020radon, he2021downsampled}. Others introduce differentiable modules, such as FBP, link between the data and image domains for end-to-end training \cite{hu2020hybrid, zhang2021clear, tao2021learning}. More advanced frameworks implement parallel or coupled dual-domain networks that simultaneously denoise projection data, recover missing values, and refine reconstructed images \cite{adler2018learn, lin2019dudonet, wang2022idol, wang2021dan}.

Despite these advancements, a fundamental challenge remains in clinical practice: the difficulty of obtaining paired training data. In supervised learning settings, high-quality ground truth images are often required, but acquiring them is costly, labor-intensive, and in many cases infeasible due
to patient privacy concerns. Existing unsupervised learning frameworks, such as CycleGAN \cite{gu2021cyclegan} and self-supervised learning \cite{niu2022noise, yu2021metal}, offer partial solutions and have shown effectiveness in several scenarios. However, their generalizability
remains limited, making them difficult to deploy reliably in many clinical scenarios.

Recently, the denoising diffusion probabilistic model (DDPM) has emerged as a contemporary generative model \cite{ho2020denoising}. Its impressive performance in natural image synthesis
has attracted significant attention in the research community \cite{saharia2022image, dhariwal2021diffusion}. Building on this success, several studies were reported on the application of DDPM to CT imaging tasks, yielding promising results in image denoising and artifact
reduction \cite{karageorgos2024denoising, xia2022low, xia2022patch}. Specifically, we incorporated data fidelity constraints into the diffusion sampling process, enabling unsupervised
reconstruction guided by a learned diffusion prior \cite{xia2023diffusion}. While diffusion models offer encouraging generative capabilities, recent studies have shown that flow-based models, especially the Poisson flow generative model (PFGM) and its generalized version (PFGM++) \cite{xu2022poisson, xu2023pfgm++} provide improved stability and sampling efficiency. 

Motivated by these breakthroughs, in this paper we extend our prior diffusion-based reconstruction framework into a Poisson flow-based model, termed Flow-Oriented Reconstruction Conditioning Engin (FORCE). Our proposed framework unifies the strengths of data fidelity and deep generative prior
with a solid theoretical foundation. Through extensive experiments, we demonstrate that our proposed FORCE achieves superior performance in terms of reconstruction accuracy and stability across various challenging CT scenarios, including low-dose imaging, sparse-view reconstruction, and metal artifact
reduction (MAR) as compared to competing unsupervised methods, making it a highly desirable meta-solution for real-world clinical applications.

\begin{figure}[t]
	\centering
	\includegraphics[width=1.0\linewidth]{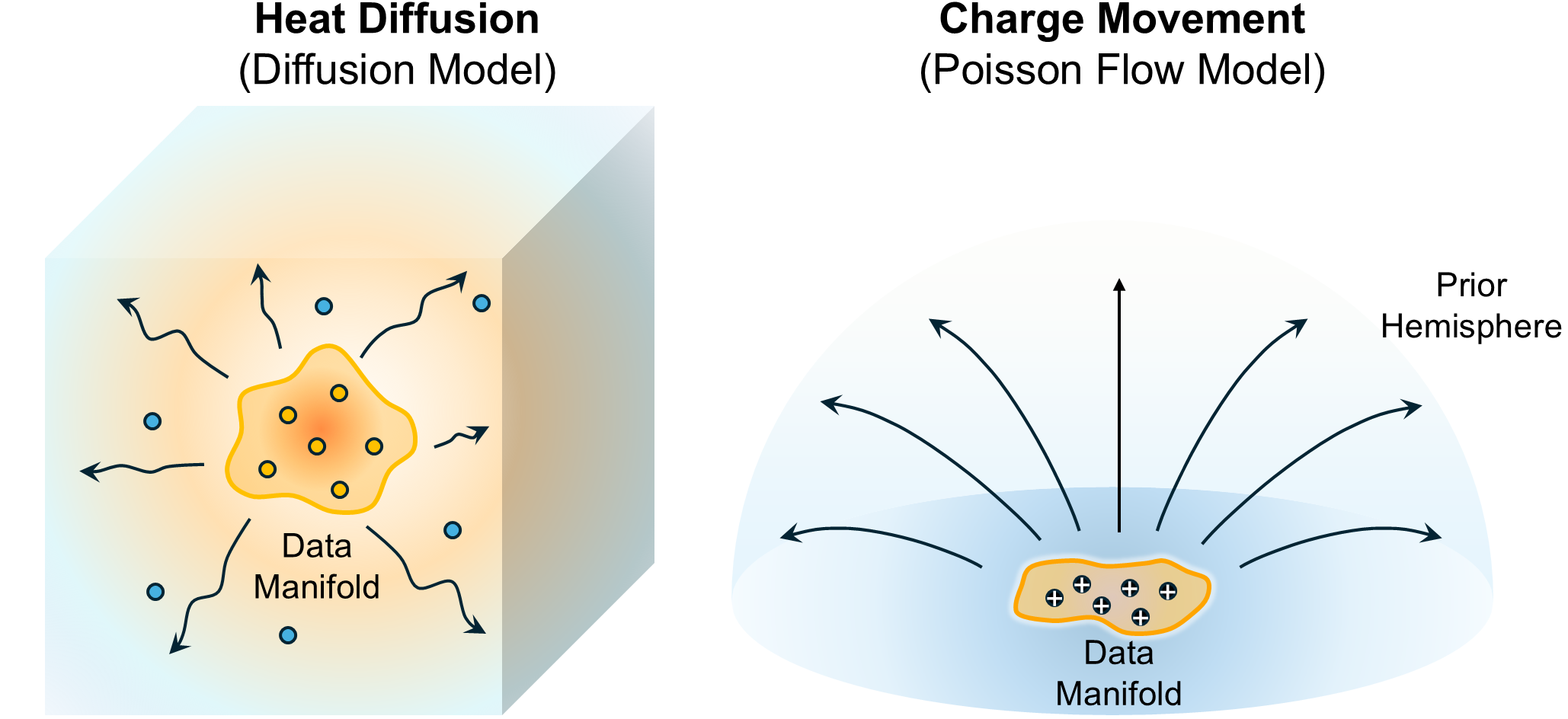}
	\caption{Conceptual comparison between diffusion and Poisson flow models. The diffusion model (left) simulates heat diffusion where noise is gradually added to the data manifold. In contrast, the Poisson flow model (right) treats data samples as charged particles defining electrical force lines away from the data manifold toward a uniform prior distribution on a hypersphere.}
	\label{fig:1}
\end{figure}

\section{Methodology}
\label{sec:2}
\subsection {Diffusion Model and Poisson Flow Model}
Both the DDPM and PFGM/PFGM++ consist of two complementary processes: a forward process that gradually corrupts training data, and a reverse process that recovers data from noise under the
same distribution of the training data. The forward process simulates a degradation mechanism, such as thermal diffusion according to thermal dynamics or motion along force lines of a physical/virtual force, while the reverse process learns to undo this corruption, enabling data generation. During training, the forward process creates paired noisy-clean samples and trains a network to estimate the underlying score function, allowing the model to perform its inference via a denoising process. As illustrated in Fig. \ref{fig:1}, diffusion and Poisson flow models differ in their physical interpretations and resultant trajectories.

In the diffusion model, the forward process describes the heat diffusion, which gradually dissipates
image features to noise. This degradation can be mathematically formulated as a stochastic differential equation (SDE) \cite{song2020score}:
\begin{equation}
    \mathrm{d}\mathbf{x} = f(\mathbf{x}, t) \, \mathrm{d}t + g(t) \, \mathrm{d}\mathbf{w},
    \label{eq:1}
\end{equation}
where $\mathbf{x}\in \mathbb{R}^N$ denotes the data, $f(\mathbf{x}, t)$ is the drift term for a deterministic trend, $g(t)$ controls the magnitude of stochastic perturbation, and $\mathrm{d}\mathbf{w}$ is the standard Wiener process (Brownian motion). The forward process transforms the data distribution into an isotropic Gaussian. Then, the reverse process samples the data distribution to synthesize realistic samples.

In contrast, PFGM/PFGM++ describes data evolution under the influence of a high-dimensional electrical field. In other words, data points are treated as charged particles that repel away from the data manifold due to the electrical field induced by the data distribution. The forward process in PFGM/PFGM++ allows the motion of particles from the data manifold to a spherically uniform prior at infinity, following the direction of the Poisson field. It is governed by the following ordinary differential equation (ODE) \cite{xu2023pfgm++}:
\begin{equation}
    \mathrm{d}\tilde{\mathbf{x}} = \mathbf{E}(\tilde{\mathbf{x}}) \, \mathrm{d}t,
    \label{eq:2}
\end{equation}
where $\tilde{\mathbf{x}} = (\mathbf{x},\mathbf{z})$ with $\mathbf{z} \in \mathbb{R}^{D}$ is the augmented data embedded in an augmented space of dimensionality $N+D$, and $\mathbf{E}(\tilde{\mathbf{x}})$ is the electric field induced by the data distribution in terms of the Poisson kernel:
\begin{equation}
   \mathbf{E}(\tilde{\mathbf{x}}) = \frac{1}{\mathbb{S}^{N+D-1}(1)} \int \frac{\tilde{\mathbf{x}} - \tilde{\mathbf{y}}}{\|\tilde{\mathbf{x}} - \tilde{\mathbf{y}}\|^{N+D}} p(\mathbf{y}) \, \mathrm{d}\mathbf{y}
    \label{eq:3}
\end{equation}
where $p(\mathbf{y})$ denotes the data density in $\mathbb{R}^{N}$ (embedded as ($\mathbf{y}, \mathbf{0}$) in $\mathbb{R}^{N+D}$), $\mathbb{S}^{N+D-1}$ is the surface area of the unit hemisphere in $\mathbb{R}^{N+D}$. This equation defines the Poisson field, which in the forward process pushes training data samples away from the data manifold. In the reverse process, we can
sample a point on a remote spherical surface, and the trace back along the corresponding force line to sample the data distribution.

Over DDPM and other diffusion-based generative AI models, PFGM++ offers several key advantages. Unlike the diffusion models, which are inspired by nonequilibrium thermodynamics and require small denoising steps in principle, PFGM is based on electrostatics and generates samples by following deterministic trajectories defined by a Poisson equation in an augmented space. This results in faster, more stable sampling with a clearer physical interpretation. Moreover, the extended version, PFGM++, generalizes the concept by introducing a dimensionality parameter $D$ for the augmented
space. Notably, when $D$ is sufficiently large, PFGM++ reduces to the diffusion model. Therefore, PFGM++ provides a more theoretically unified framework for superior robustness, efficiency, and sample quality.

\subsection{EDM-based PFGM++}
In EDM \cite{karras2022elucidating}, Karras et al. derived a probability flow ODE that governs the evolution of data samples under a continuous-time diffusion process:
\begin{equation}
    \mathrm{d}\mathbf{x} = -\dot{\sigma}(t) \, \sigma(t) \nabla_\mathbf{x} \log p(\mathbf{x}; \sigma(t)) \, \mathrm{d}t,
    \label{eq:4}
\end{equation}
where $\nabla_\mathbf{x} \log p(\mathbf{x}; \sigma(t))$ is the score function \cite{song2020score}, $\sigma(t)$ is the noise schedule, and $\dot{\sigma}(t)$ denotes the time derivative of $\sigma(t)$. The model is trained to denoise noisy samples under the following objective:
\begin{equation}
    \mathcal{L}(\bm{\theta}) = \mathbb{E}_{\mathbf{y}\sim p(\mathbf{y})} \mathbb{E}_{\bm{\epsilon}\sim \mathcal{N}(0, \bm{I})} \|\mathbf{f}_{\bm{\theta}}(\mathbf{y}+\sigma \bm{\epsilon}; \sigma) - \mathbf{y}\|_2^2,
    \label{eq:5}
\end{equation}
where $\mathbf{f}_{\bm{\theta}}$ is the denoising network. The score function can be estimated using:
\begin{equation}
    \nabla_\mathbf{x} \log p(\mathbf{x}; \sigma) = (\mathbf{f}_{\bm{\theta}}(\mathbf{x}; \sigma) - \mathbf{x})/\sigma^2.
    \label{eq:6}
\end{equation}

Let $r(\tilde{\mathbf{x}}) = \|\mathbf{z}\|_2$ denote the norm of the augmented data point $\tilde{\mathbf{x}}$. Then, this data point lies on a hypersphere of radius $r$. In PFGM++, computing the derivative with respect to the radius $\mathrm{d}\mathbf{x}/ \mathrm{d}r$ is often more convenient than using $\mathrm{d}\mathbf{x}/ \mathrm{d}t$ with the EDM formulation. Since $\|\bm{\epsilon}\|_2\sim \sqrt{D}$ for $\bm{\epsilon}\sim\mathcal{N}(0, \bm{I}_D)$, it is natural to set $r = \sigma \sqrt{D}$ to ensure consistency between the EDM noise scale and the PFGM radial coordinate. This alignment allows the model to adapt a EDM-type conditioned supervision for the hyperspherical structure of PFGM++. With this mapping, the PFGM++ training process can be formulated as
\begin{equation}
    \mathcal{L}(\bm{\theta}) = \mathbb{E}_{\mathbf{y}\sim p(\mathbf{y})} \mathbb{E}_{\mathbf{v}\sim \mathcal{U}(\mathbb{S}^{N+D-1})} \|\mathbf{f}_{\bm{\theta}}(\mathbf{y}+r \mathbf{v}; \sigma) - \mathbf{y}\|_2^2,
    \label{eq:7}
\end{equation}
where $\|\mathbf{v}\|_2=1$ is a unit vector sampled uniformly on the hypersphere $\mathbb{S}^{N+D-1}$, generated via $\mathbf{v}=\mathbf{u}/\|\mathbf{u}\|_2$ with $\mathbf{u} \sim\mathcal{N}(0,\bm{I}_{N+D})$. Algorithm \ref{alg:1} outlines the training procedure for PFGM++. During training, a scaling coefficient is sampled from a Beta distribution and used to perturb the norm of
the directional vector. The transformation arises from the fact that the ratio of squared norms of independent Gaussian vectors follows a Beta distribution, aligning with the geometric decomposition of energy across orthogonal components in high dimensions. Such perturbation improves the numerical
stability of training and enhances the model’s generalizability. Overall, the algorithm unifies EDM-type supervision with Poisson-driven hyperspherical sampling, enabling a geometrically grounded yet data-efficient training paradigm.

\begin{algorithm}[t]
\caption{PFGM++ Training with EDM}
\label{alg:1}
\KwIn{Dataset $p(\mathbf{y})$, noise schedule \( p(\sigma) \)}
\KwOut{Trained model $\mathbf{f}_{\bm{\theta}}$}
\While {\textnormal{not converged}}{
    \(\mathbf{y} \sim p(\mathbf{y})\), \( \sigma \sim p(\sigma) \) \\
    \( r = \sigma  \sqrt{D} \), \( \beta \sim \text{Beta} \), \( R = r\sqrt{\beta/(1 - \beta)} \) \\
    \( \mathbf{u} \sim \mathcal{N}(0, \bm{I}) \), \( \mathbf{v} = \mathbf{u} / \|\mathbf{u}\|_2 \) \\
    \(\mathcal{L}(\bm{\theta}) =  \|\mathbf{f}_{\bm{\theta}}(\mathbf{y}+R\mathbf{v}; \sigma) - \mathbf{y}\|_2^2\) \\
    Update \( \bm{\theta} \)
    }
\end{algorithm}

\subsection{Foundation Model FORCE for CT Reconstruction}
Model-based iterative reconstruction (MBIR) for CT originates from the Bayesian framework, where the posterior likelihood is maximized as:
\begin{equation}
	\max_{\mathbf{x}} \log{p(\mathbf{x},\mathbf{p})}=\log{p(\mathbf{p} \mid \mathbf{x})} + \log{p(\mathbf{x})}.
	\label{eq:8}
\end{equation}
where $\mathbf{x}$ denotes the reconstructed image and $\mathbf{p}$ is the measured projection data. Based on this maximum a posteriori (MAP) formulation, the classical MBIR model can be written as:
\begin{equation}
    \min_{\mathbf{x}} \frac{1}{2} \|H\mathbf{x}-\mathbf{p}\|_2^2 + \lambda R(\mathbf{x}),
    \label{eq:9}
\end{equation}
where $H$ is the system matrix representing the forward projection operator, $R(\cdot)$ is the regularization term reflecting prior knowledge, and $\lambda$ controls the trade-off between data fidelity and regularization. The choice of $R(\mathbf{x})$ is critical and traditionally hand-crafted, with popular choices including total variation (TV), nonlocal self-similarity, low-rank priors, and others. 

Importantly, the regularization term $R(\mathbf{x})$ implicitly encodes a prior distribution of images in a relevant domain. In this context, the score function in DDPM, which estimates the
gradient of the log-prior distribution, can be viewed as a learned counterpart to hand-crafted regularizers. Thus, the iterative MBIR process can be naturally linked to the sampling process in the DDPM inference. By introducing the data fidelity constraint into the diffusion process, we transform unconditional generative sampling into a posterior-guided denoising and artifact removal procedure.

Several prior studies explored this connection. For example, Song et al. \cite{song2022solve} introduced sinogram-domain value substitution to address sparse-view CT reconstruction. Chung et al. \cite{chung2022improve} applied conditioning via gradient descent to enforce data consistency. In our previous papers, we employed the SART algorithm as a conditioning mechanism to address both lowdose and sparse-view CT scenarios. In this paper, we extend this idea further by embedding the data fidelity in the prior learning with PFGM++. This leads to a unified conditioning framework applicable to a variety of CT modes/tasks, including low-dose imaging, sparse-view reconstruction, and metal artifact reduction, without requiring any paired training data. The sampling pipeline introduced in the proposed FORCE model is detailed in Algorithm \ref{alg:2}.

\begin{algorithm}[t]
\caption{FORCE Sampling for CT Reconstruction}
\label{alg:2}
\KwIn{Corrupted projection data \(\mathbf{p}\), time schedule \(\{t_i\}_{i=0}^{T}\)}
\KwOut{Reconstructed image \(\mathbf{x}_T\)}
Initialize with a low-quality image \(\mathbf{x}_{\text{init}}\) from \(\mathbf{p}\) \\
\(\mathbf{u} \sim \mathcal{N}(0, \bm{I})\), \(\mathbf{v} = \mathbf{u} / \|\mathbf{u}\|_2\) \\
\(\sigma_0 = \sigma(t_0)\), \(r_0 = \sigma_0 \sqrt{D}\) \\
Initialize sampling: \(\mathbf{x}_0 = \mathbf{x}_{\text{init}} + r_0 \cdot \mathbf{v}\)\\ 
\(\xi_0 = 1\) \\

\For{\(i = 0\) \KwTo \(T - 1\)}{
    \(\xi_i = \frac{1 + \sqrt{1 + 4\xi_{i-1}^2}}{2}\) \\
    
    \(\bar{\mathbf{x}}_i = \textnormal{Conditioning}(\mathbf{x}_i, \mathbf{p})\) \\

    \(
    \hat{\mathbf{x}}_i = \arg\min_{\mathbf{z}} \frac{1}{2} \|\mathbf{z} - \mathbf{f}_\theta(\bar{\mathbf{x}}_i; \sigma_i)\|_2^2 + \lambda \cdot \textnormal{TV}(\mathbf{z})
    \)

    \(
    \hat{\mathbf{x}}_i \leftarrow \hat{\mathbf{x}}_i + \frac{\xi_{i-1} - 1}{\xi_i}(\hat{\mathbf{x}}_i - \hat{\mathbf{x}}_{i-1})
    \)
    
    \(\mathbf{d}_i = (\mathbf{x}_i - \hat{\mathbf{x}}_i) / \sigma_i\) \\
    
    \(\mathbf{x}_{i+1} = \mathbf{x}_i + (t_{i+1} - t_i) \mathbf{d}_i\)
}
\end{algorithm}

\begin{figure}[t]
	\centering
	\includegraphics[width=0.8\linewidth]{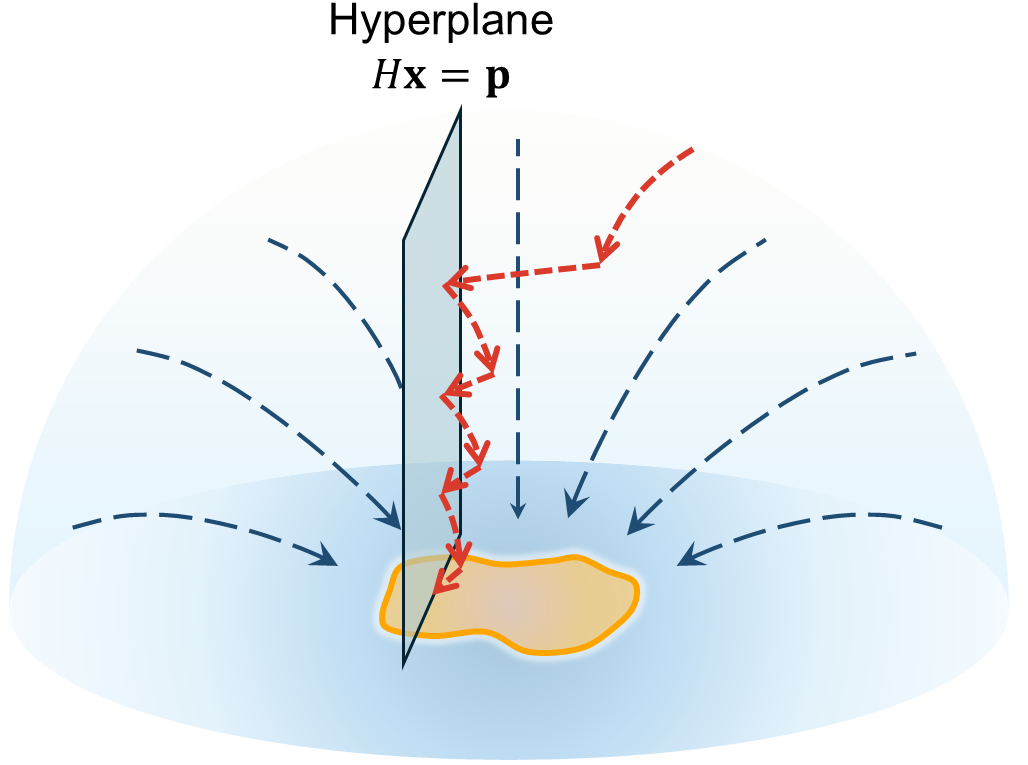}
	\caption{Conditioned sampling trajectories. The red lines illustrate the reverse Poisson flow aided by data fidelity constraints, represented by the observation hyperplane \( H\mathbf{x}=\mathbf{p} \). The trajectory gradually converges to the solution at the intersection of the feasible domain and the data manifold.}
	\label{fig:2}
\end{figure}

\paragraph{Conditioning Methods}
Conditioning refers to incorporating a relevant fidelity constraint and other appropriate priors into the generative sampling process to reconstruct a desirable tomographic image while suppressing hallucinated structures. In CT reconstruction, the observation model $H\mathbf{x}=\mathbf{p}$ defines a feasible set as a hyperplane in the image space. As shown in Fig. \ref{fig:2}, effective conditioning guides the sampling trajectory toward the intersection of the data manifold and this feasible region. The choice of a conditioning strategy should be tailored to the characteristics of the specific imaging mode/task and the reliability of the data.

For \textbf{low-dose CT}, where projection data are heavily contaminated with noise, aggressive conditioning may lead to premature convergence to noisy reconstructions. In this case, we recommend using the Regularization by Denoising (RED) method \cite{romano2017little}:
\begin{equation}
    \bar{\mathbf{x}}_i = \arg\min_{\mathbf{x}}  \frac{1}{2}\|H\mathbf{x}-\mathbf{p}\|_2^2 + \frac{\eta}{2}\mathbf{x}^T(\mathbf{x}-\mathbf{x}_i),
    \label{eq:10}
\end{equation}
which is a quadratic optimization problem and can be efficiently solved using the conjugate gradient (CG) algorithm. This approach balances the data consistency with the proximity to the current iterate.

For \textbf{sparse-view CT}, projection data are often assumed to be clean, with relatively low noise. In such cases, data fidelity can be strictly enforced. A common method is the Projection Onto Convex Sets (POCS) framework, which alternates between projections onto the feasible domain and the learned data manifold. A more efficient alternative is Ordered Subsets SART (OS-SART) \cite{wang2004ordered}, which typically converges toward the feasible set after a single pass per iteration.

In \textbf{metal artifact reduction (MAR)} tasks, projection data outside the metal trace are highly reliable, while the regions inside the trace contain corrupted values or data voids. In this setting, an effective strategy is to replace the unreliable regions during conditioning while preserving the accurate parts of the data. Following the method proposed in \cite{zhang2018convolutional}, we generate a prior image from the current estimate $\mathbf{x}_i$, forward-project it, and substitute metal-affected regions in the sinogram with synthetic data patches. This ensures data consistency while allowing the Poisson field to inpaint artifact-affected regions in a principled manner.

\paragraph{Total Variation and Momentum}
We incorporate both
total variation (TV) regularization and momentum into the FORCE sampling process. Given the characteristics of CT images, PFGM++ may generate high-frequency noise or minor structural artifacts during sampling. TV regularization is effective at suppressing such noise and artifacts, thereby enhancing the signal-to-noise ratio and improving overall image quality.

Momentum, a second-order optimization technique, is widely adopted in iterative reconstruction and variational inference. Introducing momentum into the sampling process allows the model to traverse the data manifold more efficiently, enabling higher image quality with fewer sampling steps. Since PFGM++ is governed by an ODE rather than an SDE, it is inherently more efficient than diffusion models and capable of producing high-fidelity reconstructions in relatively few iterations. Naturally, momentum provides an additional performance boost without increasing algorithmic complexity signficantly.

It is suggested that both TV regularization and momentum acceleration are applied to the intermediate variable $\hat{\mathbf{x}}_i$, which represents a denoised estimate of the clean image at each step. We do not recommend applying the momentum directly to $\mathbf{x}_i$, the current sample in the generative trajectory, as the noise levels between successive iterations can vary significantly. Estimating the momentum from these noisy samples may introduce instability or amplify artifacts.

\begin{figure}[t]
	\centering
	\includegraphics[width=0.8\linewidth]{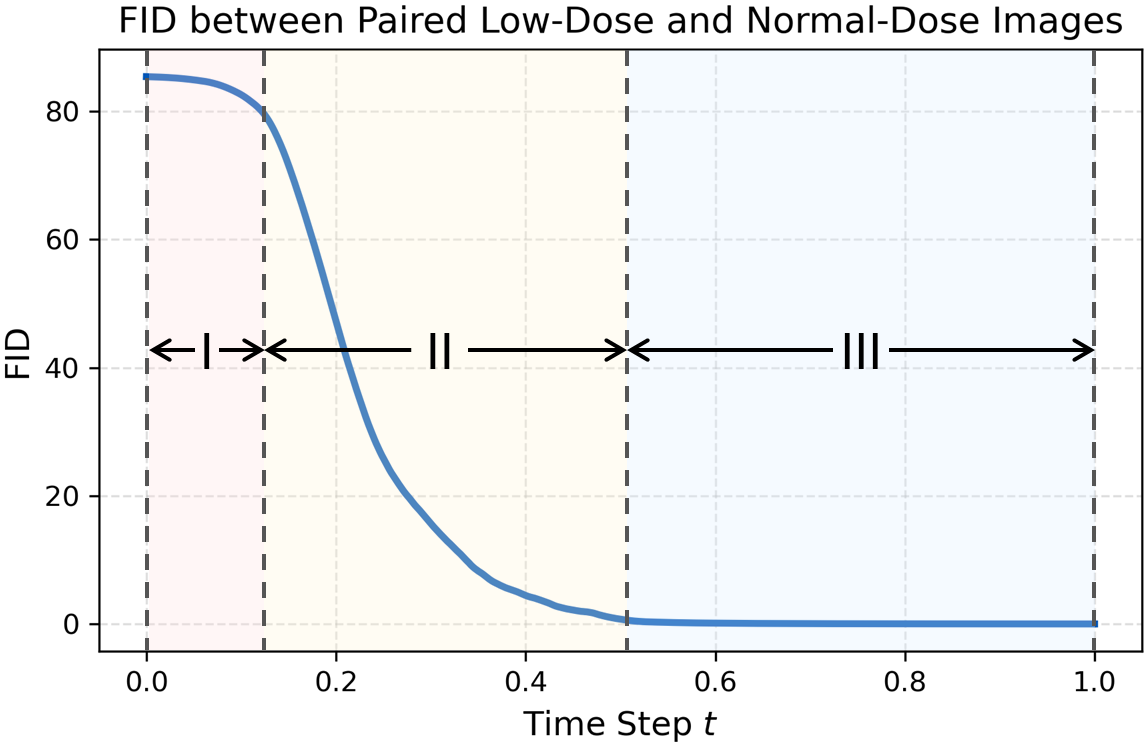}
	\caption{FID between paired low-dose and normal-dose CT images under increasing noise levels. The curve shows three phases: (I) artifact-dominant, (II) alignment, and (III) noise-dominant. Phase II is ideal for sampling, balancing artifact suppression and structural restoration.}
	\label{fig:3}
\end{figure}

\begin{figure*}[t]
	\centering
	\includegraphics[width=1.0\linewidth]{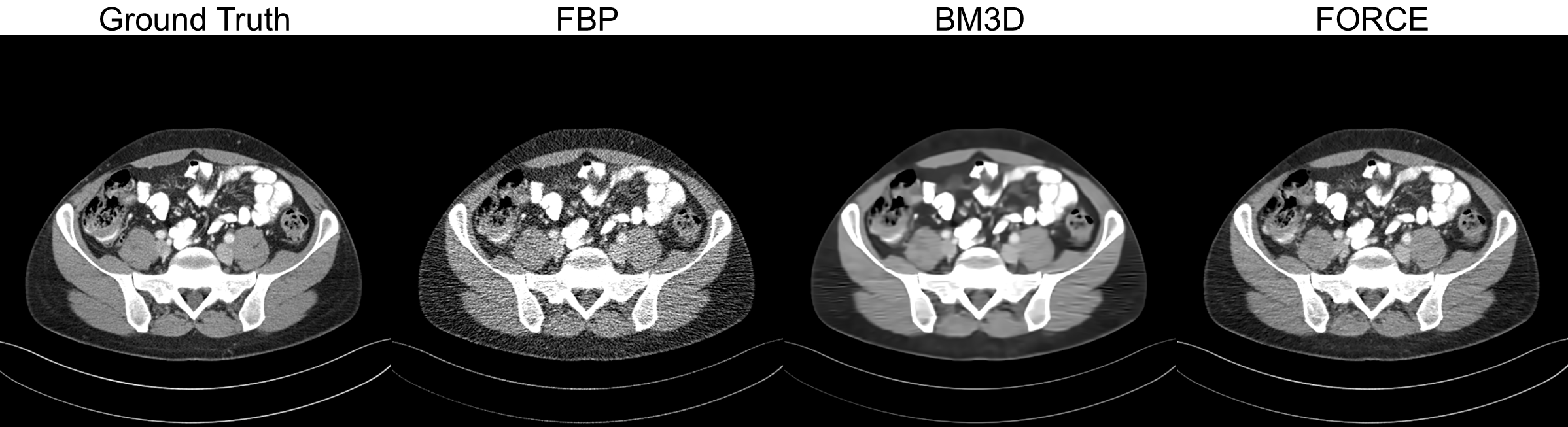}
	\caption{Pelvis reconstructions with different methods from 25\% dose data. The display window is [-160, 240] HU.}
	\label{fig:4}
\end{figure*}

\begin{figure*}[t]
	\centering
	\includegraphics[width=1.0\linewidth]{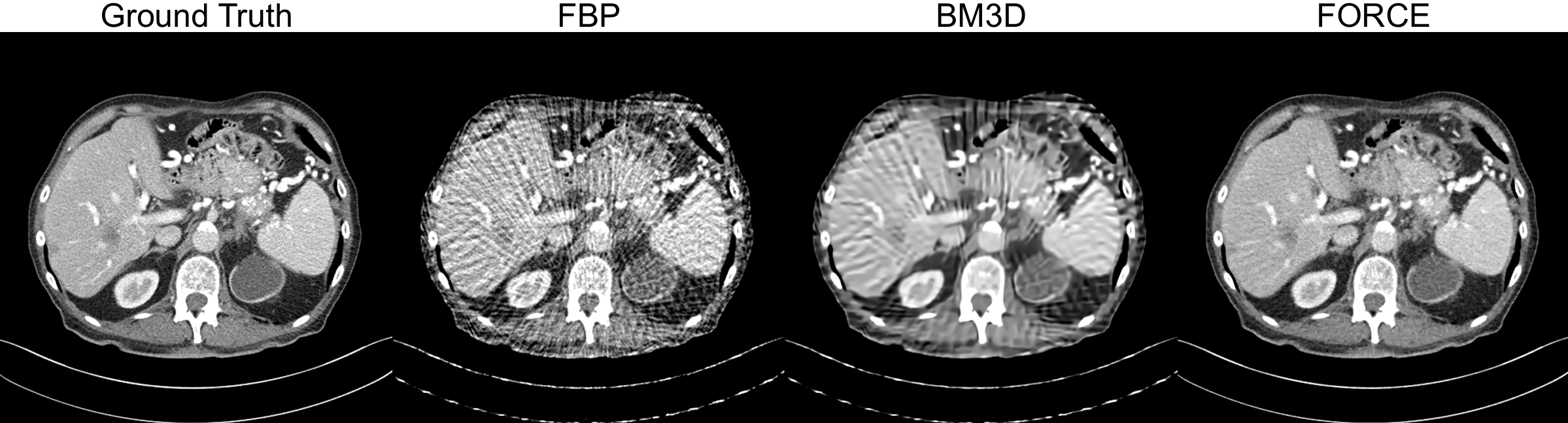}
	\caption{Abdominal reconstructions with different methods from 96 projection views. The display window is [-160, 240] HU.}
	\label{fig:5}
\end{figure*}

\begin{figure*}[t]
	\centering
	\includegraphics[width=1.0\linewidth]{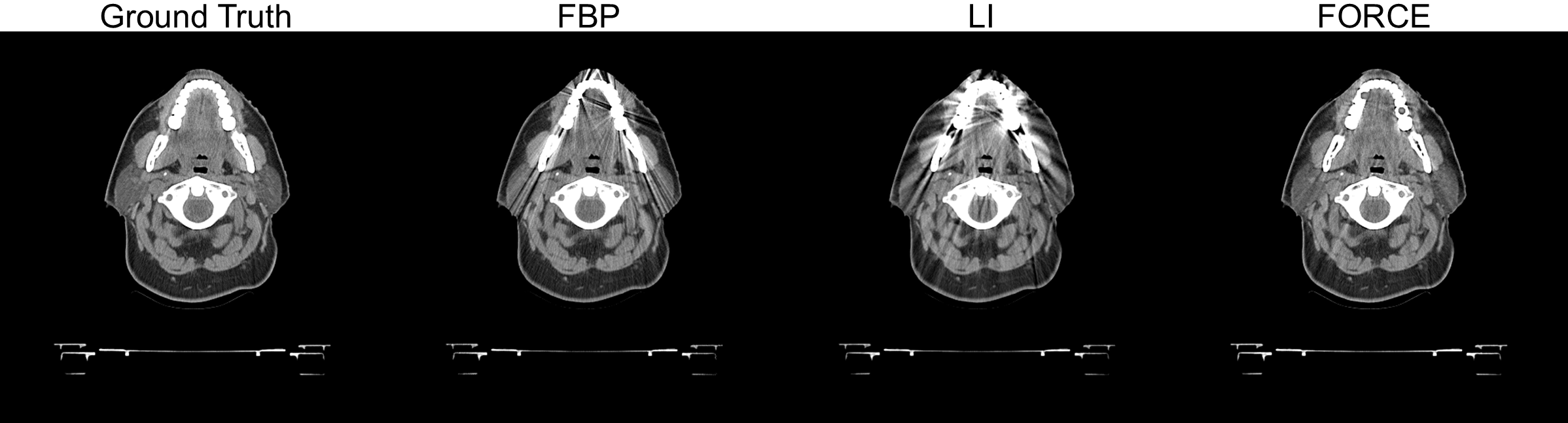}
	\caption{Dental reconstruction with different methods from metal corrupted data. The display windows is [-160 240] HU.}
	\label{fig:6}
\end{figure*}

\paragraph{Starting Sampling Time}
In conventional image generation tasks, sampling typically begins from pure Gaussian noise to encourage diverse generalization. However, CT reconstruction has a fundamentally different goal: recovering a precise image from corrupted observations, where both noise and artifacts are present in the input. In this context, starting the sampling process from pure noise is unnecessary and
potentially counterproductive. Instead, it is more appropriate to initialize from a mid-level noisy point that balances artifact suppression and structural restoration.

Let $\mathbf{x}$ and $\mathbf{y}$ denote a pair of low-dose and normal-dose CT images, respectively. Their corresponding distributions $p(\mathbf{x})$ and $p(\mathbf{y})$ differ due to the presence of artifacts and noise in $\mathbf{x}$. As noise is progressively added through the diffusion process, i.e., sampling from $p(\mathbf{x} + \sigma_t\bm{\epsilon})$ and $p(\mathbf{y} + \sigma_t\bm{\epsilon})$ with increasing $t$, the two distributions gradually converge. Eventually, at sufficiently high noise levels, the content of the original images becomes negligible and the distributions become indistinguishable.

To investigate this phenomenon, we selected paired low-dose and normal-dose CT images and measured the Fr\'{e}chet Inception Distance (FID) between them after injecting varying levels of noise. As shown in Fig. \ref{fig:3}, the resulting FID curve can be segmented into three distinct phases:
\begin{itemize}
    \item \textbf{Phase I – Artifact Dominant}: At low noise levels, the inherent noise and artifacts in the low-dose image still significantly influence the distribution. Starting sampling from this region risks retaining unwanted corruption, as the model may interpret noise patterns as meaningful structure.
    \item \textbf{Phase II – Alignment Zone}: With moderate noise, the added perturbation begins to suppress low-dose-specific artifacts, and the distributions of $\mathbf{x}$ and $\mathbf{y}$ rapidly align. This region presents an optimal trade-off—residual noise is sufficiently overwhelmed, yet structural information remains to guide sampling.
    \item \textbf{Phase III – Noise Dominant}: At high noise levels, both distributions converge to near-identical isotropic Gaussian noise. However, the underlying anatomical structure is almost entirely lost, potentially leading to inefficient sampling and slower convergence.
\end{itemize}
We posit that the optimal initial sampling time should be chosen from Phase II. At this point, added noise effectively masks the artifacts present in the input, preventing them from biasing the reconstruction. Meanwhile, enough structural signal is retained to accelerate the denoising trajectory and improve sampling efficiency.

\section{Experiments and Results}
\label{sec:3}
To evaluate the effectiveness of our FORCE model in solving different CT reconstruction problems, we conducted experiments on two public benchmark datasets: ”the 2016 NIH-AAPM-Mayo Clinic Low-Dose CT Grand Challenge” dataset and ”the AAPM CT Metal Artifact Reduction (CTMAR) Grand Challenge” dataset.

The AAPM-Mayo dataset provides 2,378 full-dose CT images with a 3 mm slice thickness from 10 distinct patients. Of these, 1,923 slices from 8 patients were used for training, while the remaining 455 slices from the other 2 patients were reserved for testing. The CT-MAR dataset comprises 14,000 slices, including 12,374 chest/abdomen/pelvis images and 1,626 head and dental images.

To simulate real-world conditions, we further processed the AAPM dataset to generate two specific CT reconstruction scenarios: (1) low-dose CT with 25\% of the original photon counts and (2) sparse-view CT with only 96 projections. Both settings were used for testing.

\begin{table}[htbp]
\centering
\caption{Quantitative comparison of different methods on low-dose CT reconstruction.}
\begin{tabular}{lccccc}
\toprule
\textbf{Metric} & \textbf{LDCT} & \textbf{BM3D} & \textbf{CycleGAN} & \textbf{Noise2Sim} & \textbf{FORCE} \\
\midrule
PSNR ($\uparrow$) & 32.25  & 35.43  & 36.17  & 36.17  & \textbf{37.32} \\
SSIM ($\uparrow$) & 0.730  & 0.894  & 0.920  & \textbf{0.928}  & 0.923 \\
FID  ($\downarrow$) & 122.86 & 100.31 & 53.92  & 76.08  & \textbf{48.58} \\
\bottomrule
\end{tabular}
\label{tab:1}
\end{table}

\subsection{Low-Dose Study}
Fig. \ref{fig:4} presents a representative pelvis CT slice reconstructed from 25\% dose data. It is evident that the BM3D method, while being effective for image noise suppression, leads to over-smoothing and blurring of fine anatomical structures. In contrast, our proposed FORCE framework successfully removed noise while preserving critical details, demonstrating its ability of balancing denoising and structural fidelity under photon-starved conditions.

\begin{table}[htbp]
\centering
\caption{Quantitative comparison of different methods on sparse-view CT reconstruction.}
\begin{tabular}{lccccc}
\toprule
\textbf{Metric} & \textbf{FBP} & \textbf{BM3D} & \textbf{CycleGAN} & \textbf{\makecell{\textbf{Sparsier2}\\\textbf{Sparse}}} & \textbf{PFGM} \\
\midrule
PSNR ↑  & 25.70 & 29.85 & 34.13 & 26.92 & \textbf{39.23} \\
SSIM ↑  & 0.479 & 0.741 & 0.877 & 0.682 & \textbf{0.947} \\
FID ↓   & 228.18 & 212.38 & 99.87 & 233.86 & \textbf{48.09} \\
\bottomrule
\end{tabular}
\label{tab:2}
\end{table}

\subsection{Sparse-View Study}
Fig. \ref{fig:5} shows an abdominal CT slice reconstructed from sparse-view data collected at only 96 projection angles. The streaking artifacts introduced by angular sparsity are global and spatially correlated, which poses a significant challenge for local or patch-based post-processing methods such as BM3D. As seen in the figure, BM3D failed to suppress these artifacts effectively. In comparison, FORCE largely reduced the artifacts and preserved anatomical boundaries such as those
around the liver and bowel, enabling improved diagnostic quality.

\subsection{Metal Artifact Study}
Fig. \ref{fig:6} illustrates a dental CT image corrupted by metal artifacts arising from high-density implants. The popular linear interpolation (LI) method was highly sensitive to the metal
trace and surrounding anatomy. As a result, LI not only failed to eliminate the artifacts but often amplified them, particularly in the regions near the teeth and jawbone. FORCE, on the other hand, effectively suppressed metal artifacts while preserving adjacent structural features. The restored image retains clinically important details of the tooth morphology, which are essential for accurate diagnosis and treatment planning.

\section{Discussion and Conclusion}
\label{sec:4}
In this study, we have proposed FORCE, a unified Poisson flow-based generative framework for CT reconstruction under various data fidelity conditions and image prior assumptions, including low-dose acquisition, sparse-view sampling, and metal artifact corruption. Unlike conventional reconstruction methods that rely on hand-crafted priors or supervised deep learning, FORCE leverages a learned data-driven prior from the Poisson flow and integrates data fidelity and other prior
constraints to guide the reconstruction process in an unsupervised manner.

Through experiments on public benchmark datasets, FORCE has demonstrated superior performance across a range of imaging modes/tasks. FORCE effectively suppresses noise and artifacts while maintaining structural fidelity, outperforming competing methods.

Despite these promising results, several directions are promising for further development to maximize FORCE's clinical utility and scientific impact:

\textbf{1. Extension to More Tomographic Tasks}
Currently, FORCE has been validated on three common CT reconstruction tasks. Future work will expand its capability to handle more clinically significant scenarios such as limited-angle tomography and interior tomography. These tasks pose challenges due to extreme data truncation, requiring FORCE to help in learning from incomplete measurements and leveraging strong priors. Through this extension, FORCE aspires to become a unified solution capable of addressing the full spectrum of CT reconstruction tasks.

\textbf{2. Scaling to 3D with Efficient Architecture}
Although the current work focuses on 2D slices, practical CT deployment requires 3D volume reconstruction. FORCE will be extended to 3D using memory-efficient strategies such as latent consistency distillation (LCD) \cite{luo2023latent} and vector-quantized autoencoders (VQ-VAEs) \cite{van2017neural}. These methods will drastically reduce training and inference costs by compressing the data into lower-dimensional latent spaces, while still preserving clinical information. This will enable FORCE to process full CT volumes using a single or limited number of high-memory GPUs.

\textbf{3. Dual-Domain and Bi-Directional Conditioning}  
FORCE naturally supports sinogram and image domain modeling. In future work, we will extend this capability to a \textit{bi-directional conditioning} framework, where models in sinogram and image domains guide each other in the reconstruction process. Inspired by the primal-dual optimization paradigm, this design enables sinogram priors to condition image sampling and vice versa, forming a tightly coupled feedback loop. This dual-domain interaction enhances both data fidelity and structural consistency, especially in cases where critical information resides across dual domains.

\textbf{4. In-Context Learning for Conditioning}
In its current form, FORCE requires manually designed conditioning strategies tailored to each specific task, which limits synergy, usability and scalability. To enhance model power and user control, in-context learning \cite{wang2023context, wang2023seggpt} will be integrated into FORCE. This innovation allows FORCE to accept prompts (e.g., textual descriptions, example images/sinograms) without retraining. These prompts can guide reconstruction to favor specific anatomical features or clinical preferences, transforming FORCE from a fixed pipeline into an interactive and adaptive reconstruction assistant. This idea builds on successful practices in segmentation and vision-language models and will make FORCE capable of generalizing across tasks and datasets with minimal engineering overhead.

\textbf{5. Foundation Model for Diverse Clinical Protocols}  
FORCE is initially envisioned as a vendor-neutral and generalizable foundation model for tomographic imaging. To achieve this goal, we will pretrain FORCE on large-scale simulated and real-world CT datasets covering multiple scanner vendors, scan protocols, and anatomical regions \cite{de2007catsim, abadi2018dukesim, jia2012gpu}. This allows FORCE to learn representations of scanner-specific characteristics and protocol-dependent variations. To support rapid adaptation, we will design the model to accept protocol-level prompts (e.g., scanner ID, scan mode, body region) as part of the input, enabling in-context modulation of the reconstruction behavior. This prompt-aware architecture not only enhances FORCE’s generalization, but also makes it possible to fine-tune the model efficiently for specific vendors and clinical applications.

In conclusion, FORCE provides a powerful and flexible solution for CT reconstruction under adverse conditions. By integrating a principled flow-field generative prior with physics-driven constraints, the proposed CT reconstruction foundation model achieves an optimal balance between data fidelity
and image prior. The approach has a strong potential for improvement and deployment in the real world for low-dose CT, CT metal artifact reduction, and many other scenarios.

\bibliographystyle{IEEEtran} 
\bibliography{ref.bib}

\end{document}